\documentclass[preprint]{revtex4}
\usepackage{graphicx}% Include figure files
\usepackage{dcolumn}% Align table columns on decimal point
\usepackage{bm}% bold math
\begin{document}
\setcounter{page}{1}
\title
{Lorentz Contracted Proton}
\author
{D. Bedoya Fierro, N. G. Kelkar and M. Nowakowski}
\affiliation
{ Dept. de Fisica, Universidad de los Andes,
Cra.1E No.18A-10, Santafe de Bogota, Colombia}
\begin{abstract}

The proton charge and magnetization density distributions can be related 
to the well known Sachs electromagnetic form factors  
$G_{E,M}({\bm q}^{2})$ through 
Fourier transforms, only in the Breit frame. The Breit 
frame however moves with relativistic velocities in the Lab and a Lorentz boost 
must be applied to the form factors before extracting the static properties of 
the proton from the corresponding densities. Apart from this, the Fourier transform 
relating the densities and form factors is inherently a non-relativistic expression. 
We show that the relativistic corrections to it can be obtained 
by extending the standard Breit equation to 
higher orders in its $1/c^2$ expansion. 
We find that the inclusion of the above corrections reduces the size of the proton 
determined from electron proton scattering data. Indeed the central value of 
the latest proton radius 
of $r_p = 0.879$ fm as determined from e-p scattering changes to 
$r_p = 0.8404$ fm after applying corrections.  
\end{abstract}
%\pacs{13.40.Gp, 14.20.Dh, 03.70.+k}
%\keywords{Electromagnetic form factors, Breit equation}
\maketitle

\section{Introduction}
Measurements of the ground state properties of the most basic element
of the constituents of stable matter, namely, the proton, have 
intrigued physicists since the sixties until now. 
The structure of the proton in fact plays an important role in atomic physics 
where experiments have reached very high precision. Finite size effects (FSE) 
due to
the proton structure can be theoretically included using different methods, with one 
of them being the Breit equation \cite{Breit,sanctis,Breit2} 
which is a typical example of 
how one derives coordinate potentials from Quantum Field Theory 
\cite{pseudoscalar,grifols,cosmology}. Using such a 
method where one evaluates the elastic electron-proton amplitude expanded in powers 
of $1/c^2$, corrections to the energy levels of the hydrogen atom due to the finite 
size of the proton have been evaluated \cite{wewithF1,wewithF2}. 
The FSE are included through the 
elastic electromagnetic form factors obtained 
from electron proton scattering cross sections \cite{bosted}.
The electromagnetic form factors as such are an 
essential part of the description of the properties of the nucleon as they 
incorporate the probability for a nucleon to absorb a virtual photon of 
four momentum squared $q^2 (= (q^0)^2 -({\bm q})^2 )$ and probe its interior. 
In the non-relativistic 
limit, the Fourier transforms of the form factors in the Breit frame 
(defined by $q^0 = 0$) 
describe the charge distribution 
$\rho_{\!_C}({\bf r})$ and magnetization current distribution 
$\rho_{\!_M}({\bf r})$ in 
the nucleon respectively \cite{reviews,arring}. An experimental determination of 
the form factors (and hence the proton charge density distribution) 
from electron proton scattering can thus enable one to determine 
the charge radius of the proton. 
On the other hand, the unprecedented precision 
of the experimental results in the hydrogen atom also allows one to probe 
the static properties of one of the components of the hydrogen atom, 
namely, the proton. The size of the proton for 
example, can be extracted from precise measurements of the difference in the energy 
levels or Lamb shifts in the hydrogen atom. 
Such an extraction performed on the 
muonic hydrogen atom led to the surprising
finding that the extracted value of $r_p = 0.84184(67)$ fm was
much smaller than the world average CODATA value of $0.8768(69)$ fm
\cite{codata6}. This so-called ``proton puzzle" was later reinforced \cite{antogpohl}
with the precise value of $r_p = 0.84087(39)$ fm obtained from muonic hydrogen 
spectroscopy. Apart from some determinations 
from standard hydrogen atom spectroscopy, the CODATA value 
largely depends on the extraction of the radius from 
electron proton scattering experiments. 
The shrunk proton gave rise to explanations ranging 
from the charge density being poorly constrained by data \cite{sick} to those 
involving large extra dimensions and non-identical protons \cite{terry}. 

In an attempt to resolve the discrepancy between the proton radius from muonic 
hydrogen spectroscopy and electron proton scattering data, we re-examine 
the connection of the 
electromagnetic form factors to the nucleon properties. 
We present a new approach to relate the form factors in momentum 
space to their coordinate space counter parts (the charge and 
magnetization densities) through a Fourier transform of the type 
$ \rho_{\!_C}({\bf r}) = e  
\int e^{ i {\bm q}\cdot {\bf r}} \,\rho_{\!_C}({\bm q}^2)\, 
d^3q/(2 \pi)^3$. The standard non-relativistic expression is obtained when 
$ \rho_{\!_C}({\bm q}^2) = \,G_E({\bm q}^2)$,  
where $G_E({\bm q}^2)$ is the well known Sachs form factor \cite{sachs}. 
The relativistic corrections for $ \rho_{\!_C}({\bf r})$ are incorporated by 
evaluating $\rho_{\!_C}({\bm q}^2)$ in the form 
$\rho_{\!_C}({\bm q}^2) = G_E({\bm q}^2) [ 1 + {\rm terms}{1 \over c^2} + {\rm terms} 
{1 \over c^4} + ...] $ using the higher order Breit equation which we derive in 
this work. Since such a relation is still valid only in the Breit frame (i.e. $q^0 =0$), 
a Lorentz boost must be applied to $\rho_{\!_C}({\bm q}^2)$ 
before extracting the mean radius from 
$r^2_p=\int \rho_{\!_C}({\bf r})r^2 d{\bf r}$. We find that the inclusion of 
the two effects (Lorentz boost and use of the higher order Breit equation) 
brings the radius extracted from electron proton scattering quite close 
to that determined from the muonic hydrogen Lamb shift \cite{antogpohl}, 
thus partly resolving the proton radius puzzle. 

\section{Theoretical approach}
The relativistic corrections to the non-relativistic $ \rho_{\!_C}({\bf r})$ 
are obtained by extending the standard Breit equation 
\cite{wewithF1,wewithF2} 
(which involves an expansion of the amplitude to order 1/c$^2$) 
\cite{LLbook,sanctis} to higher orders. 
The proton electric potential $V_p({\bf r})$ in this equation is used to find 
the density $\rho_{\!_C}({\bf r})$ via the Poisson equation, 
$\nabla^2 V_p = - 4 \pi \rho_{\!_C}$. 
The hyperfine interaction terms in the Breit equation are shown to be 
related to the magnetization density $\rho_{\!_M}({\bf r})$. 
In what follows, we shall see that an interesting  
outcome of the calculation is that 
the charge form factor $ \rho_{\!_C}({\bm q}^2)$ 
appearing in the Fourier transform, depends on the magnetic 
form factor $G_M({\bm q}^2)$ and $\rho_{\!_M}({\bm q}^2)$ appearing in 
the Fourier transform of the magnetization density, $\rho_{\!_M}({\bf r})$, 
depends on $G_E({\bm q}^2)$. 

\subsection{Potentials and densities}
In order to make the approach clear let us begin with 
the standard Breit equation for the Hamiltonian H$_B$ \cite{wewithF1} which 
results from the $1/c^2$ expansion 
of the elastic electron proton transition matrix element M$_{fi}$. 
This amplitude can be written as, 
M$_{fi} = w^{\dagger}_{S^{\prime}_e} \, w^{\dagger}_{S^{\prime}_p} \, 
\hat{{\rm H}}_B ({\bf p}_e, {\bf p}_p; \mbox{\boldmath $\sigma$}_e, 
\mbox{\boldmath $\sigma$}_p; {\bm q}) \, 
w_{S_e} \, w_{S_p}$, where $w_{S_e,S_p}$ are two component spinors. In the diagonal 
case, $S^{\prime}_e = S_e$, $S^{\prime}_p = S_p$ and we write 
M$_{fi} = M_{fi}({\bf p}_e, {\bf p}_p; {\bm \xi}_e, {\bm \xi}_p; {\bm q})$, 
where we used $w^{\dagger}_S \mbox{\boldmath $\sigma$} w_S = {\rm Tr}[\rho 
\mbox{\boldmath $\sigma$}] 
= {\bm \xi}$ with $\rho$ being the spin density matrix. Let us now rearrange terms 
from the Breit Hamiltonian H$_B$ such that, H$_B = e V_p({\bm q}) \, + \, {\bf \mu}_e 
\cdot {\bf B}({\bm q}) \, + ....$, where, ${\bf \mu}_e = - (e/2m_e) 
\mbox{\boldmath $\sigma$}_e$. 
$V_p({\bm q})$ is the potential part remaining after separating all the 
$\mbox{\boldmath $\sigma$}_i$ 
operator dependent and differential operator ${\bf p}_i$ dependent terms. 
Here $i$ is either $e$ or $p$. 
In addition, 
we choose $V_p({\bm q})$ not to contain the electron mass as the 
electric proton potential should not depend on the probe. 
These restrictions allow 
$V_p({\bm q})$ to be interpreted as a proton electric potential in momentum 
space. For the 
standard Breit equation at lowest order in $1/c^2$, with form factors, 
this indeed leads to \cite{wewithF2}, 
\begin{equation}\label{protonpot}
V_p({\bm q}) = 4 \pi e \,\biggl [ F_1\biggl
(\frac{1}{{\bm q}^{2}} \biggr ) -  F_2\biggl
(\frac{1}{4m_p^2c^2}\biggr)\biggr]\, 
 = 4 \pi e \, 
\biggl [ {G_E({\bm q}^{2})\over {\bm q}^{2}} \biggr ] . 
\end{equation}
where $e$ is the positive charge of the proton. 
The Fourier transform of $V_p({\bm q})$ is then the electric potential 
$V_p({\bf r}) = 4 \pi e \, 
\int e^{ i {\bm q}\cdot {\bf r}} \,(G_E({\bm q}^2)/ {\bm q}^2)\, d^3q/(2 \pi )^3$. 
The Laplacian of $V_p({\bf r})$, namely,
$\nabla^2 V_p({\bf r}) = - 4 \pi e \, 
\int e^{i {\bm q}\cdot {\bf r}} 
\,G_E({\bm q}^2)\, d^3q/(2 \pi)^3$ taken together with   
$\nabla^2 V_p({\bf r}) = - 4 \pi  \rho_{\!_C}({\bf r})$ then brings us to the 
standard definition of the proton charge density 
$ \rho_{\!_C}({\bf r}) = e\, 
\int e^{i {\bm q}\cdot {\bf r}} \,G_E({\bm q}^2)\, d^3q/(2 \pi)^3$. 
Applying similar restrictions to the magnetic field in the 
second term in H$_B$, i.e., the magnetic field of the proton $B({\bm q})$ should 
not contain any electron mass or operator ${\bf p}_i$ dependence, the terms which 
remain (apart from the Coulomb term) are those corresponding to the 
hyperfine interaction. The hyperfine interaction potential 
with form factors \cite{wepramana} is given as, 
\begin{eqnarray}\label{potqhfs}
V({\bm q})_{hfs}  &=&  \alpha \biggl [ \, 
{(\mbox{\boldmath $\sigma$}_e \cdot 
{\bm \sigma}_p) \over 4 m_e\, m_p c^2}  - 
{(\mbox{\boldmath $\sigma$}_e \cdot {\bm q}) \cdot (
\mbox{\boldmath $\sigma$}_p \cdot {\bm q}) \over 
4 m_e\, m_p c^2  {\bm q}^2} \, \biggr ] 
G_M({\bm q}^2) \nonumber \\
&=&\, {\bf \mu}_e \cdot {\bf B}({\bm q})\, ,
\end{eqnarray}
with ${\bf \mu}_e = - (e/2m_e) \mbox{\boldmath $\sigma$}_e$ as defined earlier. 
The magnetic field of the proton is thus, 
\begin{equation}\label{magp}
{\bf B}({\bm q}) =  e \,\biggl [ \, 
{{\bm q} 
(\mbox{\boldmath $\sigma$}_p \cdot {\bm q}) - 
{\mbox{\boldmath $\sigma$}_p {\bm q}^2}  \over  
2 m_p c^2  {\bm q}^2} \, \biggr ] G_M({\bm q}^2)\, .
\end{equation}
Taking the Fourier transform of ${\bf B}({\bm q})$ and using the static Maxwell 
equation ${\bf \nabla} \times {\bf B}({\bf r}) = 4 \pi {\bf j}({\bf r})$, we can identify 
\begin{equation}
{\bf j}({\bf r}) = {e \over 4 \pi}\,\int \, {d^3q \over (2 \pi)^3} 
\, e^{i {\bm q}\cdot {\bf r}}
{G_M({\bm q}^2) \, (i{\bm q} \times \mbox{\boldmath $\sigma$}) 
\over 2 m_p c^2}\, , 
\end{equation}
thus implying ${\bf j}({\bf r}) \propto {\bf \nabla} \times {\bf M}$ with 
${\bf M} = \rho_{\!_M}({\bf r}) \mbox{\boldmath $\sigma$}_p$ which defines the proton magnetization 
distribution $\rho_{\!_M}({\bf r}) = 
\int \,e^{i {\bm q} \cdot {\bf r}}\, 
G_M({\bm q}^2) \, d^3q/(2 \pi)^3$. In the diagonal case, by the replacement of 
${\bm \sigma}$ by the polarization vector ${\bm \xi}$, we can conclude that 
polarized protons will have a magnetic field of the form, 
${\bf B}({\bf r}) \propto ({\bm \xi} \cdot {\bm \nabla}) {\bm \nabla} 
\tilde {\rho}_{\!_M}\, -\, {\bm \xi} \rho_{\!_M}$, where, 
$\tilde {\rho}_{\!_M}\, = \int e^{i {\bm q}\cdot {\bf r}} 
\,(G_M({\bm q}^2)/{\bm q}^2)\, d^3q/(2 \pi)^3$. 
All the conclusions drawn above and derived at the lowest order in the 
relativistic expansion are, of course, valid if we include relativistic corrections. 

\subsection{Higher order Breit equation}
The procedure to obtain the 
Breit potential at higher orders 
using the electron proton scattering amplitude  
is exactly the same as that described in 
\cite{wewithF1, wewithF2} except that the proton and electron wave 
functions which were written in \cite{wewithF1,wewithF2} using the non-relativistic 
approximation with corrections up to order $1/c^2$ are now replaced by those 
containing relativistic corrections up to order $1/c^6$. This is done by 
using the 
Foldy Wouthuysen transformation \cite{foldy}, $\Psi_{\!_{FW}} = U \Psi_{\!_D}$, where  
\begin{equation}\label{foldy1}
U = \, \sqrt{(E + m c^2) \over 2 E}
\biggl ( 
1 \, +\, {\beta \mbox{\boldmath $\alpha$} \cdot {\bf p}c \over E + mc^2} 
\biggr )\, ,
\end{equation}
$H_{\!_D} \Psi_{\!_D} = E \Psi_{\!_D}$, 
$E = \sqrt{{\bf p}^2 c^2 + m^2 c^4}$, $H_{\!_{FW}} \Psi_{\!_{FW}} 
= \beta E \Psi_{\!_{FW}}$ and ${\bm \alpha}$, $\beta$ the usual Dirac 
matrices.  
%\begin{equation}
%{\bm \alpha} = 
%\left(\begin{array}{c} 0 \, \, \,\,{\bm \sigma} \\
% {\bm \sigma} \, \, \, \,0 \end{array}\right) \, ,\, \,
%\beta = 
%\left(\begin{array}{c} 1 \, \, \, \,\, \,\,\, \,0 \\
% 0\, \, \, -1 \end{array}\right) \, .
%\end{equation}
It then follows that \cite{silenko}  
$\Psi_{\!_{FW}} = [E(1 + \beta) / \sqrt{2E (E + mc^2)}] \Psi_{\!_D}$, 
where, $\Psi_{\!_{FW}}$ 
contains both the positive and negative energy solutions. The upper
and lower components $\Psi_{\!_{FW}}^+$ and $\Psi_{\!_{FW}}^-$ of 
$\Psi_{\!_{FW}}$ can be shown to be 
related to the Dirac upper and lower components $\phi_{\!_D}$ and $\chi_{\!_D}$ 
respectively as \cite{silenko}
\begin{equation}\label{foldy2}
\Psi^+_{\!_{FW}} = \, \sqrt{2E \over E+mc^2} 
\left(\begin{array}{c} \phi_{\!_D} \\
 0 \end{array}\right) \, ,
\Psi^-_{\!_{FW}} = \, \sqrt{2E \over E+mc^2} 
\left(\begin{array}{c} 0 \\
\chi_{\!_D} \end{array}\right) 
\end{equation}
The relativistic energy $E$ of the particle includes also its rest 
energy $m c^2$ which must be excluded in arriving at a non-relativistic approximation. 
We must therefore replace $\Psi$ (FW or D) by $\Psi^{\prime}$ defined as 
$\Psi = \Psi^{\prime} \,e^{-i mc^2 t / \hbar}$. This leads to a  
relation between the upper and lower components $\phi$ and $\chi$ 
of $\Psi^{\prime}$ \cite{LLbook} which is given by, 
\begin{equation}\label{spinoreq}
\chi = {1 \over 2 m c}\,\, \biggl [ 
1 \, +\, {E_S\over 2 m c^2} \biggr ]^{-1} \, 
\mbox{\boldmath $\sigma$} \cdot {\bf p}\, \phi,
\end{equation}
where $E_S$ is the energy eigenvalue in the Schr\"odinger equation.
Identifying the upper component 
$\Psi_{\!_{FW}}^{\prime +}$ of $\Psi_{\!_{FW}}$ 
with the non-relativistic Schr\"odinger spinor 
$w$, we get, $w \, \sqrt{(E+mc^2)/2E} = \phi$. Finally, expanding 
$E = (p^2 c^2 + m^2 c^4)^{1/2}$ and replacing for $\phi$ in terms of $w$ 
in $\chi$, we obtain the spinor to be used in the calculation of the 
amplitude $M_{fi} = e^2 (\bar{u}^{\prime}_e \Gamma_e^{\mu} u_e) \, 
D_{\mu \nu} (q^2)\,(\bar{u}^{\prime}_p \Gamma_p^{\nu} u_p)$ as 
\begin{equation}\label{spinoreq2}
u_i =\sqrt{2m_i}\left(\begin{array}{c}(1-{p_i^2\over 8m_i^2c^2}+
\frac{\lambda_1p_i^4}{m_i^4c^4}+\frac{\lambda_3p_i^6}{m_i^6c^6}) \, w_i\\
(1-\frac{\lambda_2p_i^2}{m_i^2c^2}+\frac{\lambda_4p_i^4}{m_i^4c^4})
{{\bm \sigma}_i \cdot {\bf p}_i \over 2m_ic}\, w_i 
\end{array}\right) \, ,
\end{equation}
with $i = e, p$ and $\lambda_1=11/128$, $\lambda_2=3/8$, $\lambda_3=-69/1024$ and 
$\lambda_4=31/128$. 
The above spinor should be contrasted with 
\begin{equation}\label{spinoreq3}
u_i =\sqrt{2m_i}\left(\begin{array}{c}(1-{p_i^2\over 8m_i^2c^2}) \, w_i\\
{{\bm \sigma}_i \cdot {\bf p}_i \over 2m_ic}\, w_i 
\end{array}\right) \, ,
\end{equation}
used to obtain the standard Breit equation \cite{LLbook}. 
Using Eq. (\ref{spinoreq2}) 
and the vertices $\Gamma_p^{\nu} = F_1^p \gamma^{\mu} 
+ \sigma^{\mu \nu} (q_{\nu}/ 2 m_p c) F_2^p$ and $\Gamma_e^{\mu} = \gamma^{\mu}$, the 
amplitude $M_{fi}$ and hence the Breit equation with form factors 
is evaluated just as in \cite{wewithF1, wewithF2}. Note that the energy transfer 
at the vertices is chosen to be zero, i.e., $q^2 = \omega^2/c^2 - {\bm q}^2$ is replaced 
by $q^2 = - {\bm q}^2$. Formally, this is achieved by going to the Breit frame. 
This is in keeping with the quasistatic approach wherein we 
are going to relate the proton potential obtained from the Breit equation to the charge 
density via the Poisson equation. The 
higher order Breit equation with form factors thus obtained is very  
lengthy and will be given elsewhere. The present work deals with 
the parts relevant for obtaining the relativistic corrections to the 
charge and magnetization densities. 

The proton electric potential $\tilde{V}_p({\bm q})$ 
with relativistic corrections  
is obtained from the higher order Breit equation in the same manner as 
explained before for the standard Breit equation.  
Dropping all terms involving the spin and momentum operators
as well as those containing the electron mass, what remains in the higher 
order Breit equation is 
%\begin{widetext}
\begin{eqnarray}\label{newpot1}
\tilde{V}_p({\bm q})&=& 4 \pi e {G_E({\bm q}^{2})\over {\bm q}^{2}} \, 
\biggl \{ 
1 - {{\bm q}^2 \over 8 m_p^2 c^2} + {3 \over 128} {{\bm q}^4 \over 
m_p^4 c^4} - {13 \over 1024} {{\bm q}^6 \over m_p^6 c^6}   \\ \nonumber
&+&{G_M({\bm q}^2) \over G_E({\bm q}^2)}{{\bm q}^4 \over 16 m_p^4 c^4} 
\biggl [1 - {7 \over 8} {{\bm q}^2 \over m_p^2 c^2} + {87 \over 128} 
{{\bm q}^4 \over  m_p^4 c^4} \biggr ] 
\biggr \} . 
\end{eqnarray} 
%\end{widetext}
The above equation can be rewritten as  
$\tilde{V}_p({\bm q}) = 4 \pi e \rho_{\!_C}({\bm q}^2)/ {\bm q}^2$, such that the 
proton electric potential, $\tilde{V}_p({\bf r}) = 4 \pi e \, 
 \int e^{i {\bm q}\cdot {\bf r}} \,(\rho_{\!_C}({\bm q}^2)/ {\bm q}^2)\, d^3q/(2 \pi)^3$ 
and 
$\nabla^2 \tilde{V}_p({\bf r}) =  - 4 \pi e \, 
\int e^{ i {\bm q}\cdot {\bf r}} \,\rho_{\!_C}({\bm q}^2)\, d^3q/(2 \pi)^3 =  - 4 \pi  
\rho_{\!_C}({\bf r})$. 
The magnetic form factor $\rho_{\!_M}({\bm q}^2)$ including 
corrections is obtained by 
examining the hyperfine 
interaction terms as mentioned before, however, in the higher order 
Breit equation. 
Noting that the terms of order $1/c^6$ and higher are of decreasing 
importance and due to the alternating sign in (\ref{newpot1}), 
the first four terms in the curly bracket in (\ref{newpot1}) can be approximated 
as $[1 + ({\bm q}^2/4m_p^2 c^2)]^{-1/2}$. 
The expressions for $\rho_{\!_C,\!_M}({\bm q}^2)$ can thus 
be summarized in an expansion effectively as,   
\begin{widetext}
\begin{eqnarray}\label{finalresult}
\rho_{\!_C}({\bm q}^2)
\simeq {G}_E({\bm q}^2) \biggl ( 1 + {{\bm q}^2 \over 4 m_p^2 c^2} 
\biggr )^{-1/2} \, +\, {G_M({\bm q}^2) {\bm q}^4 
\over 16 m_p^4 c^4} \biggl (
1 + {a {\bm q}^2 \over 4 m_p^2 c^2} \biggr )^{-b} 
\\ \nonumber
\rho_{\!_M}({\bm q}^2)
\simeq {G}_M({\bm q}^2) \biggl ( 1 + {{\bm q}^2 \over 4 m_p^2 c^2} 
\biggr )^{-1/2} \, -\, {G_E({\bm q}^2) {\bm q}^2 \over 4 m_p^2 c^2} \biggl (
1 + {a {\bm q}^2 \over 4 m_p^2 c^2} \biggr )^{-b} \, ,
\end{eqnarray}
\end{widetext}
with $a = 19/7$ and $b = 49/38$. It is interesting that 
$\rho_{\!_C}({\bm q}^2)$ and $\rho_{\!_M}({\bm q}^2)$
depend on both the 
$G_E$ and $G_M$ Sachs form factors and have relativistic corrections of a 
similar form with the same exponents $a$ and $b$. 
Note also that the exponent $-1/2$ in the first terms is approximate 
(in contrast to the exact $[1 + ({\bm q}^2/4m_p^2 c^2)]^{-1/2}$ 
in \cite{rosenfeld, foldyyen, friaradv, friar2}). 
%since a Taylor expansion of $[1 + ({\bm q}^2/4m_p^2)]^{-1/2} = 
%1 - {\bm q}^2/8m_p^2 \, +\, 3\bm q^4/128 m_p^4 \, -\, 5 {\bm q^6}/1024 m_p^6 \,% + ....$. 
At order $1/c^2$, the expression for $\rho_{\!_C}({\bm q}^2) \simeq G_E({\bm q}^2) 
(1 - {\bm q}^{\, 2}/8m_p^2 c^2)$ is independent of 
$G_M$ as in \cite{rosenfeld,foldyyen, friaradv, friar2, yenny}, however, 
the magnetic form factor at order $1/c^2$ reduces to 
$\rho_{\!_M}({\bm q}^2) \simeq 
G_M({\bm q}^2) (1 - {\bm q}^{2}/8m_p^2 c^2) - 
G_E({\bm q}^{\, 2}){\bm q}^{2}/
4 m_p^2 c^2$ and contains apart from the Darwin term ${\bm q}^2/8 m_p^2$, a 
term dependent on $G_E$. 

\subsection{Lorentz boost}
Since we chose the energy transfer in the evaluation of the electron - proton 
scattering amplitude, $\omega = 0$, the above form factors are similar to those 
usually given in the so-called Breit frame. 
An additional important relativistic correction arises 
due to the Lorentz contraction of the spatial distributions in the 
Breit frame \cite{kelly}. 
The latter has been discussed at length in the first reference of 
\cite{kelly} where the 
author proposes the use of the Fourier transform of 
$G_{E,M}^L({\bm q}^{2}) = G_{E,M}({\bm q}^2) [ 1 + 
({\bm q}^{2}/4m_p^2)]^{\lambda_{E,M}}$, rather than the Fourier transform of 
$G_{E,M}({\bm q}^2)$ in order to determine the density distributions of the nucleon. 
With $\lambda_{E,M}$ being model dependent constants, they 
eventually appear as parameters in the determination of the proton radius 
and other moments. 
The author in the first reference in \cite{kelly} 
fitted the form factor data to obtain $\lambda_{\!_E} = \lambda_{\!_M}= 2$ in 
agreement with some \cite{mitrakumari} while in contrast with other predictions 
\cite{otherlam} of $\lambda_{\!_E} = 0$ and $\lambda_{\!_E} = \lambda_{\!_M} = 1$ 
based on soliton and cluster models.

\section{Corrected radii and fourth moments}
The standard way of defining the $n^{th}$ moment of the charge and magnetization 
distribution in literature \cite{bern} follows from a consideration of the Fourier 
transforms of the Sachs form factors in the Breit frame. 
It makes sense to dwell a little bit on the basics of the definition of the second moment, i.e., 
\begin{equation}\label{newx0}
<r^2> = \int \, r^2 \, \rho ({\bf r}) \, d^3 r\, .
\end{equation}
Starting with the 
Fourier transform of $G({\bm q^2})$, namely, 
$G({\bm q}^2) = \int e^{-i\vec{q} \cdot \vec{r}} \rho({\bf r}) d^3r/(2\pi)^3$ 
we can readily show that
\begin{eqnarray} \label{newx1}
G({\bm q}^2) &=&{1 \over 2\pi^2} \int_0^{\infty} r^2 \rho(r) 
{\sin(|{\bm q}|r)\over |{\bm q}|r} \, dr \nonumber \\
 &= & {1 \over 2\pi^2} {1\over |{\bm q}|} \, 
\int_0^{\infty} r \rho(r) \biggl [ |{\bm q}|r - {|{\bm q}|^3 r^3 \over 6} \, +\, ....
\biggr ]\nonumber \\
&= & {1 \over 2 \pi^2} \, \biggl [ 
\int_0^{\infty} r^2 \rho(r) dr - {{\bm q}^2 \over 6 } \int_0^{\infty} r^4 \, 
\rho(r) dr \, + \, ...\, \biggr ]
\end{eqnarray}
leads to the standard result  
\begin{equation} \label{newx2}
 - {6 \over G(0)} {dG({\bm q}^2)\over d{\bm q}^2}\biggr|_{{\bm q}^2 =0} = 
\int r^4 \, \rho(r) dr = <r^2>.
\end{equation} 
Eq. (\ref{newx1}) is equivalent to writing  
\begin{equation}
G({\bm q}^2)/G(0) \, =\, 1 \, - \, {1\over 6} \, \langle r^2 \rangle {\bm q}^2 
\, + {1\over 120} \, \langle r^4 \rangle \, {\bm q}^4\, - \, ... \,, 
\end{equation}
where $\langle r^n \rangle$ is the $n^{th}$ moment of the electric or magnetic 
distribution. Neither (\ref{newx0}) nor (\ref{newx1}) are relativistic invariants. 
The form factor $G$ which depends on the four momentum transfer is 
an invariant and sometimes one finds in the literature the expansion
\begin{equation}\label{newx3}
G(q^2) = 1 \,+\, a q^2 \, + \, ... 
\end{equation}
with $q^2=q_{\mu}q^{\mu}$ being the four-momentum transfer. This is then followed
by an expression of the first moment proportional to $dG(q^2)/dq^2$ taken at $q^2=0$.
This in turn might lead to the confusing conclusion that the proton radius is a
Lorentz invariant. To resolve the confusion let us first note that we would get
the same result by writing $G(\omega=0, {\bm q}^2)=1-a{\bm q}^2+ ...$ and taking the
derivative with respect to ${\bm q}^2$ evaluated at ${\bm q}^2=0$ which agrees
with (\ref{newx2}) and, of course, (\ref{newx0}). Therefore, we would face 
a paradox here: by using (\ref{newx3}) and its derivative with respect to the 
four momentum squared $q^2$, 
it seems like we have found a Lorentz invariant quantity and this is equivalent to
a Lorentz non-invariant result (up to the minus sign which is absorbed into the
definition). The resolution of the paradox lies in the meaning
of the condition, $q^2=0$. With $q^2 = \omega^2 - {\bm q}^2$, 
it either means that $\omega^2={\bm q}^2 \neq 0$ 
(in which case we have a real photon) or $\omega=|{\bm q}|=0$.  
It is impossible to exchange a real photon in the
t-channel exchange diagram in elastic electron-proton scattering and 
hence we have to drop the first possibility. The second
choice is, however, equivalent to first choosing the frame ($\omega=0$ implies that we 
have chosen the Breit frame again) 
and then ${\bm q}^2=0$ is necessary to extract the Taylor coefficient (the radius).
In short, even if $dG(q^2)/dq^2$ is invariant, the condition $q^2 = 0$ makes the 
radius defined using $dG(q^2)/dq^2$ at $q^2 =0$, a Lorentz non-invariant quantity 
(as the condition forces one to choose $\omega =0$).     

%The second moment for example is determined by 
%\begin{equation}
%\langle r^2 \rangle \, = - {6 \over G(0)} {dG({\bm q}^2) \over d{\bm q}^2}\biggr |_ 
%{{\bm q}^2 =0}\, .
%\end{equation}

If $\rho_{\!_C}({\bf r})$ and 
$\rho_{\!_M}({\bf r})$ (defined by $ \rho_{\!_C}({\bf r}) =  
e \int e^{ i {\bm q}\cdot {\bf r}} \,\rho_{\!_C}({\bm q}^2)\, 
d^3q/(2 \pi)^3$ and 
$ \rho_{\!_M}({\bf r}) =  
 \int e^{ i {\bm q}\cdot {\bf r}} \,\rho_{\!_M}({\bm q}^2)\, 
d^3q/(2 \pi)^3$), 
receive relativistic corrections as given in (\ref{finalresult}), 
so will the corresponding radii.
Hence, the proton moments including the relativistic corrections and the Lorentz boost 
are defined here as:
\begin{equation}
\langle \tilde{r}_{\!_{E}}^2 \rangle^L 
= - {6 \over \rho_{\!_C}^L(0)}\,  {d\rho_{\!_C}^L \over d{\bm q}^2}\biggr |_{{\bm q}^2 =0}
\end{equation}
and 
\begin{equation}
\langle \tilde{r}_{\!_{E}}^4 \rangle^L = 
{60 \over \rho_{\!_C}^L(0)}\,  {d^2\rho_{\!_C}^L \over d({\bm q}^2)^2} 
\biggr |_{{\bm q}^2 =0}
\end{equation}
(with $\rho_{\!_C}^L({\bm q}^2) = 
\rho_{\!_C}({\bm q}^2) \, [ 1 + ({\bm q}^{2}/4m_p^2c^2)]^
{\lambda_{\!_{E}}}$) \, . 
Replacing from (\ref{finalresult}) for $\rho_{\!_C}({\bf q}^2)$ leads to 
\begin{eqnarray}
\langle\tilde{r}_{\!_E}^2 \rangle^L &=& 
\langle r_{\!_E}^2 \rangle + {3 \over 4 m_p^2 c^2} \, (1 - 2\lambda_{\!_E})\\
\langle \tilde{r}_{\!_E}^4 \rangle^L 
&=& \langle r_{\!_E}^4 \rangle - {5\over m_p^2 c^2} \langle r_{\!_E}^2 \rangle \biggl 
(\lambda_{\!_E} - {1\over 2}\biggr ) 
\nonumber \\
&+&{15 \over 4 m_p^4 c^4} (\lambda_{\!_E}^2 - 2 \lambda_{\!_E} + 2 \mu_p) +{45\over 16 m_p^4c^4}
\nonumber
\end{eqnarray}
The magnetic radius with relativistic and 
Lorentz boost corrections is given by, 
\begin{eqnarray}
\langle \tilde{r}_{\!_M}^2 \rangle^L &=& \langle r_{\!_M}^2 \rangle + {3\over 4 m_p^2 c^2}
\biggl [ 1 \, +\, 
{2\over \mu_p} - 2\lambda_M \biggr ]
\end{eqnarray}
The relativistic corrections alone (giving $\tilde{r}_p$ and $\tilde{r}^4$ in Table I) 
arising from (\ref{finalresult}) can be 
found by setting $\lambda_{E,M} = 0$.  
\begin{figure}[h]
\includegraphics[width=9cm,height=9cm]{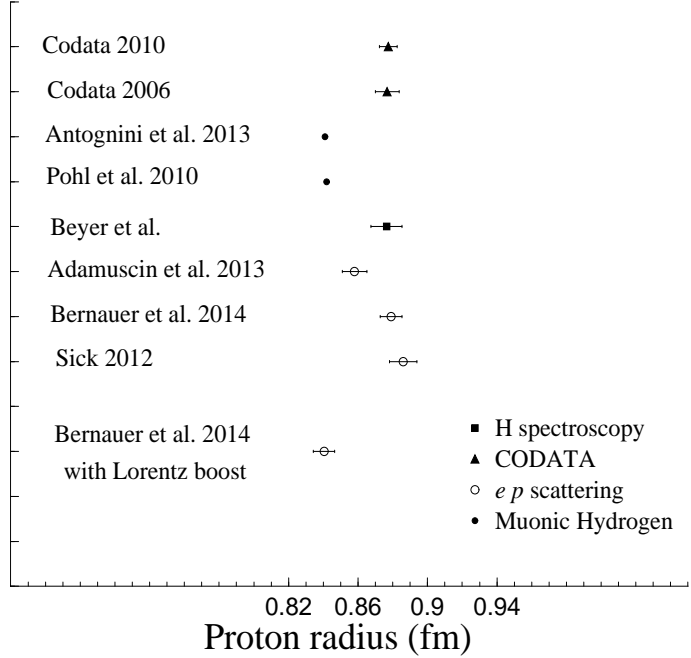}
\caption{\label{fig:eps1}
Comparison of the proton radius values extracted from
the muonic hydrogen Lamb shift \cite{antogpohl},
CODATA values \cite{codata6} and some recent analyses
\cite{sick,bern} of $ep$ scattering data.
The hydrogen spectroscopy average is from \cite{beyer}.}
\end{figure}
\begin{table}[ht]
\caption{Corrections to the proton charge radius $r_{\!_p} = \langle 
r_{\!_E}^2 \rangle^{\!^{1/2}}$ 
in fm. The fourth moments $r^4 = \langle r_{\!_E}^4 \rangle$ 
of the proton charge distribution with corrections (up to order 1/c$^2$ 
and 1/c$^4$) are given in the brackets 
(in fm$^4$). The first column gives the usual uncorrected values obtained from 
$e p$ scattering, the second column shows the increase in values due to 
relativistic corrections, the third and fourth the reduction due to Lorentz boost and
finally the last two columns display the effects of both corrections together.
}
\begin{tabular}{|l|l|l|l|l|l|l|}
  \hline
   & $r_p$ & $\tilde{r}_p$  &$r_p^L$   &$r_p^L$  
& $\tilde{r}_p^L$  & $\tilde{r}_p^L$ \\
   & ($r^4$) & ($\tilde{r}^4$)  &($r^4$)$^{\!^L}$   &($r^4$)$^{\!^L}$
& ($\tilde{r}^4$)$^{\!^L}$  & ($\tilde{r}^4$)$^{\!^L}$ \\
  
 & & &$\lambda_{\!_E} = 1$ & $\lambda_{\!_E} = 2$& $\lambda_{\!_E} = 1$ & 
$\lambda_{\!_E} = 2$\\
   \hline
  Dipole \cite{bosted} & 0.811 & 0.831 & 0.769 & 0.725 & 0.790 & 0.747 \\
    up to 1/c$^4$   & (1.083) & (1.202) & (0.937) & (0.806) & (1.049) & (0.911) \\
    up to 1/c$^2$   & (1.083) & (1.156) & (0.937) & (0.792) & (1.010) & (0.864) \\
  \hline
  \cite{alber} Fit I & 0.884 & 0.903 &0.846 &0.806  & 0.865 & {\bf 0.826} \\
 up to 1/c$^4$    & (1.788) & (1.920) & (1.615) & (1.457) & (1.740) & (1.574) \\
  up to 1/c$^2$   & (1.788) & (1.875) & (1.615) & (1.442) & (1.702) & (1.529) \\
  \hline
 \cite{alber} Fit II & 0.866 & 0.885 &0.827  &0.786 & {\bf 0.847} & 0.807 \\
  up to 1/c$^4$  & (1.623) & (1.752) & (1.457) & (1.306) & (1.579) & (1.420) \\
 up to 1/c$^2$   & (1.623) & (1.706) & (1.457) & (1.291) & (1.540) & (1.374) \\
  \hline
\cite{walch} & 0.858& 0.877 & 0.819 & 0.777& {\bf 0.839} & 0.798\\
up to 1/c$^4$   &(1.488) &(1.616) &(1.325) &(1.177) &(1.446) &(1.290) \\          
 up to 1/c$^2$  &(1.488) &(1.570) &(1.325) &(1.162) &(1.407) &(1.244) \\          
\hline
\cite{bern} & 0.8795& - &0.8404  & 0.801 & - & -\\
\hline
\end{tabular}
\end{table}

The effect of the Lorentz boost in general is to reduce the radius 
and the fourth moment of the proton charge 
as compared to that obtained from $G_E({\bm q}^{\, 2})$ in the Breit frame. 
The relativistic corrections introduced with the use of 
$\rho_{\!_C}({\bm q}^2)$ obtained from 
the higher order 
Breit potential, in general, increase the radius of the proton. 
However, a fortuitous combination of the two effects, brings the proton radius 
closer to $r_p = 0.84087(39)$ fm 
obtained from precise Lamb shift measurements \cite{antogpohl}. For 
a Lorentz boost with $\lambda = 1$ such an agreement is favoured by 
Fit II in \cite{alber} which gives $\tilde{r}_p^L = 0.847$ fm 
and $\tilde{r}_p^L = 0.839$ fm obtained from 
the bump-tail parametrization (with parameters from Table II in 
\cite{walch}). Indeed, if we apply the Lorentz boost with $\lambda =1$ to the 
central value of the radius $r_p = 0.879$ fm deduced recently 
by Bernauer et al. \cite{bern} 
we obtain $r_p^L = 0.84043$ fm which is once again close to the 
muonic hydrogen spectroscopy result \cite{antogpohl}. This is demonstrated 
in Figure 1. 
The reason for applying only the Lorentz boost and not the 
entire relativistic corrections is the following: 
Bernauer et al. include in their analysis, the ``Feshbach correction" which 
as stated above Eq.(20) in \cite{bern} is in agreement with the Coulomb 
correction of Rosenfelder \cite{rosenfeld} at ${\bm q}^2 = 0$. This correction of 
Rosenfelder is similar to the relativistic corrections of the present paper upto
order $1/c^2$ (compare Eq.(7) of \cite{rosenfeld} with 
the first term in Eq.(\ref{finalresult}) of 
the present work). It would lead to a double counting if we would apply the 
relativistic correction of our work to the radius of Bernauer et al. and hence we
apply only the Lorentz boost. 
Though we do not show explicitly, similar corrections would also 
shift the other radii in Fig.1, extracted from $e p$ scattering, to lower values. 
The proton magnetic radius,   
$r_M = 0.87$ fm \cite{epstein}, with relativistic and Lorentz boost corrections 
changes to $r_M = 0.865$ fm. 
Finally, we must emphasize that the proton is characterized fully by all its moments 
and not just the radius. The corrections in Eqs (\ref{finalresult}) introduce 
a significant change in $\langle r^4 \rangle$ too.  
%The fourth moments were shown in \cite{miller} to govern the size of the third 
%Zemach moments of the proton charge distribution. 

Finally, in passing we mention that the proton structure corrections as such are 
also dependent on the theoretical formalism used to calculate them. We refer the 
reader to Ref. \cite{wewithF1} for a detailed 
discussion of the proton structure corrections using different formalisms. 

\section{Summary}
The relations between charge/magnetization densities and the 
electromagnetic form factors are necessarily 
of a non-relativistic nature. 
In other words, relativistic corrections can be computed and 
the standard relation between the Sachs form factors ($G_E({\bm q}^2)$ and 
$G_M({\bm q}^2)$) and the densities is valid only at the lowest order 
of the non-relativistic expansion. 
To compute the relativistic 
corrections in a consistent way we employed the higher order Breit equation in which, 
for instance, terms independent of the probe, spin and momentum operators should 
correspond to the proton electric potential in momentum space. 
Using the Poisson equation, this potential gives us the 
relativistically modified charge density. A similar procedure 
can be found for the magnetization density. Both results are valid in the Breit frame. 
Hence using a Lorentz transformation suggested in the literature, we can bring them 
to the rest frame of the proton and calculate the modified moments of the 
proton charge and magnetization densities. 
An interesting outcome of the manipulations, i.e., 
including relativistic corrections and the 
Lorentz transformation is that the proton radius from 
$ e p$ scattering experiments comes 
closer to the result obtained from muonic hydrogen spectroscopy.

\appendix*
\section{Coefficients in the wave function expansion}
A free spin $1/2$ particle is described by a four component wave function 
satisfying the Dirac equation. It is however, 
often desirable to convert this equation to a two component equation of the 
Pauli type. Methods attempting to do this however encounter difficulties if one 
wishes to go beyond the lowest order in the $v/c$ expansion. A method proposed 
by Foldy and Wouthuysen \cite{foldy} however overcomes these difficulties. 
Their treatment involves a unitary transformation which block diagonalizes the 
Dirac Hamiltonian and eventually splits the Dirac equation 
into two uncoupled equations of the
Pauli type, describing particles in positive- and negative-energy states, 
respectively. Since the procedure to carry out the Foldy-Wouthuysen
transformation is given below Eq. (\ref{foldy1}) in the main text, 
here we only write the intermediate steps for obtaining
the coefficients in Eq. (\ref{spinoreq2}).

We start with $w \, \sqrt{(E+mc^2)/2E} = \phi$ as given below 
Eq. (\ref{spinoreq}) and expanding $E = (p^2 c^2 + m^2 c^4)^{1/2}$, we obtain,
\begin{equation}\label{upper1}
 \phi = \biggl [ 1 - {p^2 \over 8 m^2 c^2} + {11\over 128} {p^4 \over 
m^4 c^4} - {69 \over 1024} {p^6 \over m^6 c^6} ...\biggr ] \, w \,.
\end{equation}
This is the upper component given in Eq. (\ref{spinoreq2}).
Replacing the above $\phi$ in Eq. (\ref{spinoreq})
namely,
\begin{eqnarray}\label{spinoreqap}
\chi &=& {1 \over 2 m c}\,\, \biggl [ 
1 \, +\, {E_S\over 2 m c^2} \biggr ] ^{-1} \, 
\mbox{\boldmath $\sigma$} \cdot {\bf p}\, \phi, \nonumber \\ 
&=& {\mbox{\boldmath $\sigma$} \cdot {\bf p} \over 2 m c}
\biggl ( 1 - {E_S \over 2 m c^2} + {E_S^2 \over 4 m^2 c^4} - {E_S^3 \over 8 m^3 c^6} 
+ .... \biggr )\, \phi \, ,\nonumber
\end{eqnarray}
using the fact that $E_S \, w = \hat{H} \, w$ where,
$$\hat{H} = {p^2 \over 2m} - {p^4 \over 8 m^3 c^2}$$
and replacing accordingly for every $E_S w$ by
$[{p^2 \over 2m} - {p^4 \over 8 m^3 c^2}] w$, we get
\begin{eqnarray}\label{spinoreqap2}
\chi &=& {\mbox{\boldmath $\sigma$} \cdot {\bf p} \over 2 m c} \, 
\biggl [ 
1 \, -\, {p^2\over 4 m^2 c^2} + {p^4\over 8 m^4 c^4} - \, ...\, \biggr ]  \, 
\biggl [ 1 - {p^2 \over 8 m^2 c^2} + {11\over 128} {p^4 \over 
m^4 c^4} \,  ...\, \biggr ] \, w \nonumber \\
& = & \biggl [ 1 - {3 \over 8}{p^2 \over m^2 c^2} + {31\over 128} {p^4 \over 
m^4 c^4} \, ...\, \biggr ] \, 
{\mbox{\boldmath $\sigma$} \cdot {\bf p} \over 2 m c}\,w \,.
\end{eqnarray}
This is the lower component in Eq. (\ref{spinoreq2}). 

We investigated another method \cite{LLbook,LandB} to obtain such an expansion
and it was gratifying to find the same coefficients as above. 
If we begin again with Eq. (\ref{spinoreq}) as the starting 
point and perform an expansion for $\chi$ after noting that 
$E_S$ is the total energy with rest energy subtracted, $\chi$ can be 
rewritten as, 
\begin{equation}\label{chiagain}
\chi = c \, \mbox{\boldmath $\sigma$} \cdot {\bf p}\, 
\biggl [ {1 \over 2 m c^2} - {p^2 \over 8 m^3 c^4} + {p^4 \over 16 m^5 c^6} - ....
\biggr ]\, \phi.
\end{equation}
Now noting that the density $\rho = \Psi^* \Psi = 
|\chi|^2 + |\phi|^2$, we obtain for the density, 
\begin{equation}\label{density}
\rho = |\phi|^2 + c^2 \biggl [ ({\bf p}^* A \phi^{\dagger} ) 
\cdot \mbox{\boldmath $\sigma$}\biggr ] \, \biggl [\mbox{\boldmath $\sigma$} 
\cdot ({\bf p} A \phi) \biggr ]
\end{equation}
where $A$ is basically the operator in square brackets in (\ref{chiagain}). 
This $\rho$ obviously differs from the Schr\"odinger expression. In order to find 
the wave equation corresponding to the Schr\"odinger equation, we must replace 
$\phi$ by another function $\phi_{Sch}$, for which the 
time independent integral would be of the form $\int |\phi_{Sch}|^2 d^3x$ as it 
should be for the Schr\"odinger equation. Hence, to obtain the required 
transformation, we write the condition
\begin{equation}\label{condition}
\int |\phi_{Sch}|^2 d^3x \, =\, \int \biggl \{ |\phi|^2\,+\,  
c^2  [ ({\bf p}^* A \phi^{\dagger} ) 
\cdot \mbox{\boldmath $\sigma$} ] \,  [\mbox{\boldmath $\sigma$} 
\cdot ({\bf p} A \phi) ] \biggr \} \, d^3 x\, .
\end{equation}
Integrating the second term by parts and after some lengthy but 
straightforward algebra, we find that the following expression 
for $\phi$, satisfies the 
relation in (\ref{condition}). 
\begin{equation}\label{upper2}
\phi = \biggl [ 1 - {p^2 \over 8 m^2 c^2} + {11\over 128} {p^4 \over 
m^4 c^4} - {69 \over 1024} {p^6 \over m^6 c^6} ...\biggr ] \, \phi_{Sch} \,, 
\end{equation}
This is however the same as Eq. (\ref{upper1}) for the upper component. 
Replacing this $\phi$ in (\ref{chiagain}), obviously leads to the same lower 
component as in (\ref{spinoreqap2}).

\end{document}